# The effect of multidisciplinary collaborations on research diversification[1]

Giovanni Abramo

*Laboratory for Studies in Research Evaluation
at the Institute for System Analysis and Computer Science (IASI-CNR)
National Research Council of Italy*
ADDRESS: Istituto di Analisi dei Sistemi e Informatica, Consiglio Nazionale delle Ricerche, Via dei Taurini 19, 00185 Roma - ITALY
giovanni.abramo@uniroma2.it

Ciriaco Andrea D'Angelo

*University of Rome "Tor Vergata" - Italy and
Laboratory for Studies in Research Evaluation (IASI-CNR)*
ADDRESS: Dipartimento di Ingegneria dell'Impresa, Università degli Studi di Roma "Tor Vergata", Via del Politecnico 1, 00133 Roma - ITALY
dangelo@dii.uniroma2.it

Flavia Di Costa

*Research Value s.r.l.*
ADDRESS: Research Value, Via Michelangelo Tilli 39, 00156 Roma- ITALY
flavia.dicosta@gmail.com

**Abstract**

This work verifies whether research diversification by a scientist is in some measure related to their collaboration with multidisciplinary teams. The analysis considers the publications achieved by 5300 Italian academics in the sciences over the period 2004-2008. The findings show that a scientist's outputs resulting from research diversification are more often than not the result of collaborations with multidisciplinary teams. The effect becomes more pronounced with larger and particularly with more diversified teams. This phenomenon is observed both at the overall level and for the disciplinary macro-areas.

*Bibliometrics; co-authorship; research specialization; Italy.*

# 1. Introduction

The challenges faced by scientific researchers have become increasingly complex, making the integration of multidisciplinary competencies ever more necessary. The result has been a growing reliance on collaborations between experts from different disciplinary areas, at times from very distant cognitive spheres (Boh, Ren, Kiesler, & Bussjaeger, 2007; Wray, 2005; Cummings, & Kiesler, 2005; Clark & Llorens, 2012; Jones, Wuchty, & Uzzi, 2008; Mo, 2016).. The editorial of a recent special issue of *Nature* began with the statement that: "To solve the grand challenges facing society – energy, water, climate, food, health – scientists and social scientists must work together" (Nature News, 2015). However, research that crosses disciplinary boundaries is difficult to catalyze and structure, and to finance, evaluate and publish (Viseu, 2015). Given this, there is a need for policy measures at various levels and on the part of different stakeholders, to appropriately incentivize collaboration between scientists of mixed disciplines.

The scenario outlined is clearly one in which the scientific community faces a paradox. The continual accumulation of knowledge within each scientific area requires ever greater specialization, serving advancements within the disciplines. This induces scientists to collaborate ever more with colleagues of their own areas, for "specialized research" (Strober, 2006). The risk is then a trend towards growing segmentation of knowledge, and a reductionist approach to the solution of scientific problems. Researchers can be confined to the closed realms of their own disciplines or sub-disciplines, and fail to respond to opportunities for application of their methods and competences beyond the immediate boundaries. However, as we have seen, it is precisely the crossing of boundaries that is necessary to address problems of intrinsically multidisciplinary nature, whose resolution requires the integration of different disciplinary knowledge (Weingart, 2000; Bozeman & Corley, 2004; Katz & Martin, 1997). The literature on the theme has as yet to even converge on the taxonomic aspects of the matters at hand. The definitions of multidisciplinarity, interdisciplinarity and transdisciplinarity have been proposed by various authors (Darbellay, 2015; Klein, 2008; Strober, 2006; Stokols et al., 2003). The pattern seems to be that of distinguishing in terms of the level of integration of the disciplines involved: the term *multidisciplinarity* is used to refer to contexts of two or more disciplines, even when they are not integrated; *interdisciplinarity* is used when there is a certain degree of integration between the disciplines involved (for example in the methods, contents, perspectives), but in spite of this there are still observable differences between them. Finally, the term *transdisciplinarity* is applied only to cases in which the integration between disciplines is so advanced as to create a shared conceptual framework, and where the attempt to distinguish the individual disciplines becomes difficult.

In bibliometrics, multidisciplinarity is often measured in terms of the "diversity" of research areas in the references cited by publications. The indicators used are often borrowed from other research areas, in particular from the biodiversity measures of ecology (Mugabushaka, Kyriakou, & Papazoglou, 2016) and the concentration measures of economics (Adelman, 1969). Shimada and Suzuki (2017) develop a framework analogous to that of biodiversity, stressing that: "From the policy perspective, a standard methodology for characterization of diversity of science is needed to enable the efficient management and breeding of diverse research responsive to socio-economic demands". Other than the examination of references cited, another approach is to analyze the disciplinary specializations of the co-authors of jointly produced publications. The underlying rational is that if the research team includes different disciplines, the output must necessarily possess some degree of multidisciplinarity. Some scholars have provided operative procedures for identifying the core discipline of specialization of an author (Abramo, D'Angelo, & Di Costa, 2017a; Mizukami, Mizutani, Honda, Suzuki, & Nakano, 2017).

Abramo, D'Angelo and Di Costa (2017b) posed the question of whether the diversification of a scientist's activity produces output of greater impact, compared to the output limited to research activity in their core discipline. At the same time, they verified the existence of gender differences in the propensity to diversification (Abramo & D'Angelo, 2017a), as well as the effect of age and academic rank (Abramo, D'Angelo, & Di Costa, 2017c). Finally, they examined whether the type of collaboration chosen (international, domestic, intramural) has an influence on the researcher's propensity to diversification or specialization (Abramo, D'Angelo, & Di Costa, 2017d).

Clearly, there is an interconnection between the theme of collaboration by researchers from different disciplines and that of specialization/diversification in their activity. In this work, with the objective of providing insights on the theme, we intend to study the relation between the character of a publication (specialized or diversified, depending on whether it is within or outside the author's dominant topic of research) and the multi-sectoral nature of the team of academic researchers that produced it. It must be noted that the attribute of "specialized" or "diversified" is not a characteristic of the publications themselves. Instead, we identify the character of the work considering the prevalent SC of the author that produced it.

The hypotheses tested are as follows:

*H1: Publications that are the fruit of research diversification by one or more of their co-authors arise from collaboration within multidisciplinary teams;*

*H2: As the number of disciplines in a research team increases, there is greater probability that the relative publications will be the fruit of research diversification by one or more of the authors.*

Finally, since the greater is the number of co-authors, the greater is the probability that a publication will be of multidisciplinary type, we test a third hypothesis:

*H3: With increasing number of authors of a publication, the probability increases that it will be the fruit of research diversification for one or more of the authors.*

The analysis will be conducted at the general level, and at the macro-area level for detection for any differences between them.

The hypotheses are tested using a purely bibliometric approach, considering the 2004-2008 scientific production by over 5000 Italian professors in the sciences, as indexed in the Web of Science (WoS).

The next section of the paper describes the dataset and methodology. Section 3 presents the results of the analysis, while Section 4 offers the conclusions and authors' comments.

## 2. Methodology and dataset

For the current work we use the dataset consisting of the scientific production by professors in the sciences of all Italian universities over the period 2004-2008,[2] as indexed in the WoS. We have chosen to exclude the social sciences, because it turns out that Italian professors in these fields have a too large share of their publication output in sources that are not covered by WoS.[3] Findings then would not be robust enough. It must be noted that the faculty staff in the sciences represents above 60% of total staff, while social scientists only 20%. Collaboration natures and norms tend to be different across disciplines. For instance, humanists highly value sole-authorship, while physicians are more likely to be involved in collaborative projects. Abramo and D'Angelo (2017b) showed that in the sciences, the average number of co-authors per paper is lowest in mathematics (2.74) and highest in medical sciences (6.13).[4] To help readers have a better understanding of the collaboration behavior across disciplines, and in Italy in particular, we refer them to the work by Abramo, D'Angelo and Murgia (2013). The authors provide an exhaustive analysis of variation in research collaboration patterns across disciplines and in fields within disciplines.

Data on the faculty at each Italian university were extracted from the database on Italian university personnel, maintained by the Italian Ministry of Education, University, and Research.[5] Beginning from the raw data of the WoS, and applying a complex algorithm to reconcile the author's affiliation and disambiguation of the true identity of the authors, each publication (article, article review and conference proceeding) is then attributed to the university professors that produced it (D'Angelo, Giuffrida, & Abramo, 2011).[6]

The methodology adopted for the current study is in part reprised from Abramo, D'Angelo, and Di Costa (2017a): the dominant topic of research of a scientist is represented by the subject category (SC) including the largest share of their publications.[7] The publications falling in this SC are classified as "specialized", and the remainder as "diversified". The science discipline classification scheme used is that of the WoS, with 166 subject categories (SCs) grouped in eight macro-areas.[8]

---

[2] A five-year publication period is considered adequate to reduce the problem of paucity of publications and year-dependent fluctuations with systematic effects on results (Abramo, D'Angelo, & Cicero, 2012).

[3] The percentages of Italian social science professors (by field) who have none of their 2001-2004 outputs covered by WoS, are: Political economy, 66.2%; Economic policy, 75.0%; Finance, 69.2%; History of economic thought, 86.7%; Econometrics, 28.0%; Applied economics, 77.4%; Business administration, 96.0%; Corporate finance, 87.2%; Financial management, 100.0%; Business organisation, 81.4%; Economics of financial intermediaries, 95.3%; Economic history, 95.3%; Commodity studies, 67.9% (D'Angelo and Abramo, 2015).

[4] Mathematics and computer science, 2.74; Physics, 4.54; Chemistry, 4.83; Earth sciences, 4.03; Biology, 5.07; Medicine, 6.13; Agricultural and veterinary sciences, 4.52; Civil engineering, 2.87; Industrial and information engineering, 3.47.

[5] http://cercauniversita.cineca.it/php5/docenti/cerca.php, last accessed 14 March 2018.

[6] The harmonic average of precision and recall (F-measure) of authorships, as disambiguated by the algorithm, is around 97% (2% margin of error, 98% confidence interval).

[7] This step is conducted by: i) identifying the scientist's production over the period of interest, as indexed in the WoS; ii) associate the publications with the subject categories of the hosting journals; iii) identify the subject category with the largest share of the scientist's publications.

[8] The macro-areas are: Mathematics; Physics; Chemistry; Earth and Space Sciences; Biology; Biomedical Research; Clinical Medicine; Engineering. Our assignment of SCs to macro-areas follows a pattern previously published on the website of ISI Journal Citation Reports, but no longer available on the current Clarivate portal. There are no cases in which an SC is assigned to more than one macro-area.

The academics included in the analysis are those who present a single core discipline,[9] and whose 2004-2008 WoS indexed publications meet (for reasons of significance) the following requirements:
- there are at least five of their publications;
- their publications fall in at least two different SCs (disciplines);
- their publications are co-authored, and only with other Italian academics.

The requirement of publications only by Italian academics is necessary because these are the ones for which we are able to disambiguate the true identity of all the authors, and therefore also identify all of their publications and their core discipline.

Table 1 summarizes the bibliometric dataset, divided by macro-area: the last column shows the incidence of authorship referred to multidisciplinary publications, meaning co-authorship with scientists of a different core discipline.

Earth and Space Sciences (46.2%) and Biology (42.8%) are the macro-areas with the greatest percentage of multidisciplinary co-authorship. For all macro-areas the percentage is less than 50%. The minimum is seen in Mathematics (22.7%), which is a disciplinary area characterized by a limited number of collaborations, and having these primarily within the macro-area.

*Table 1: Dataset of the analysis*

| Macro-area | Disciplines | Professors | Publications | Authorship | Multidisciplinary |
|---|---|---|---|---|---|
| Biology | 29 | 703 | 583 | 1,284 | 550 (42.8%) |
| Biomedical Research | 14 | 539 | 503 | 1,108 | 448 (40.4%) |
| Chemistry | 8 | 691 | 699 | 1,746 | 639 (36.6%) |
| Clinical Medicine | 40 | 826 | 662 | 1,431 | 511 (35.7%) |
| Earth and Space Sciences | 12 | 248 | 239 | 485 | 224 (46.2%) |
| Engineering | 39 | 1,423 | 1,853 | 4,148 | 1,364 (32.9%) |
| Mathematics | 6 | 355 | 549 | 1,109 | 252 (22.7%) |
| Physics | 18 | 547 | 692 | 1,293 | 528 (40.8%) |
| Total | 166 | 5,332 | 4,919 | 12,604 | 4,516 (35.8%) |

Note that the attribution of a publication as specialized or diversified is not univocal, instead being relative to each author that produced it: a publication having two co-authors, for example, could be specialized for one and diversified for the other. For operative purposes, the unit of analysis is the instance of co-authorship, represented by the scientist-publication combination. As an illustration, we take the case of a work published in *Cellular Microbiology*, a journal associated with the subject categories Cell Biology and Microbiology.

> Cenci et al. (2004). A synthetic peptide as a novel anticryptococcal agent. *Cellular Microbiology*, 6(10), 953-961. DOI:10.1111/j.1462-5822.2004.00413.x

Of the eight authors, all academics, three are on faculty at the University of Parma and five at the University of Perugia (Table 2). The publication in question thus determines eight different observations in the dataset, one for each author. However, for four authors the publication results as "specialized", since among the subject categories associated with the journal there is the core discipline of that author; for the other four the publication is instead diversified, on the basis of the convention adopted. In any case, the publication in question is multidisciplinary, and in particular is one with collaboration between experts of three different disciplines (Immunology, Infectious Diseases, Microbiology).

---
[9] There could be cases of academics with two or more core fields, i.e. with publications evenly distributed among them.

*Table 2: An example of the splitting process of a publication in the dataset*

| Author | Affiliation | Core discipline | Macro-area | Diversified |
|---|---|---|---|---|
| CONTI Stefania | University of Parma | Microbiology | Biology | NO |
| MAGLIANI Valter | University of Parma | Microbiology | Biology | NO |
| POLONELLI Luciano | University of Parma | Microbiology | Biology | NO |
| BISTONI Francesco | University of Perugia | Immunology | Biomedical Research | YES |
| CENCI Elio | University of Perugia | Infectious Diseases | Biomedical Research | YES |
| MENCACCI Antonella | University of Perugia | Microbiology | Biology | NO |
| PERITO Stefano | University of Perugia | Immunology | Biomedical Research | YES |
| VECCHIARELLI Anna | University of Perugia | Immunology | Biomedical Research | YES |

## 3. Analysis and results

To test the three research hypotheses raised in the introduction, we introduce three variables:
- *Multi_discipline*, is a dummy variable taking the value of 1 when a publication is produced by co-authors having different core disciplines, 0 in the contrary case;
- *n_discipline*, is a discrete variable indicating the number of different core disciplines associated with the co-authors of a publication;
- *n_authors* is a discrete variable indicating the number of co-authors in the byline of the publication.

In the next section we present the results from two types of analyses: first the univariate tests of the hypotheses, followed by a multivariate logistic regression.

### 3.1 One-way hypothesis test

Given the dichotomous character of the variable *Multi_discipline*, we use Pearson's $\chi^2$ test. For the other two variables (*n_discipline* e *n_authors*) we use the Wilcoxon rank-sum test, known as the Mann-Whitney two-sample statistic, based on the hypothesis that two independent samples (i.e., unmatched data) are from populations with the same distribution. Table 3 presents the p-values from these tests, with parenthetical indication of the signs of percentage difference of co-authorship for the "diversified" tag: the data in the last line confirm that at the overall level, with 0.05 level of significance, the null hypothesis is rejected in favor of the alternative hypothesis for all the variables considered. Further, the data in the second column attest to a significant and positive association between the character of the author's publication (diversified or specialized) and the presence (or absence) of a multidisciplinary team in the byline for the publication itself, in all the macro-areas investigated.

In other words, independent of the author's macro-area, the diversified character of their publication seems associated with presence of co-authors specialized in core disciplines other than their own; vice versa, their specialized publications seem more frequently associated with mono-disciplinary collaborations/bylines.

This result is confirmed by the data in column 3: with increasing number of disciplines of specialization among the co-authors of a publication, there is increasing probability that the publication falls in a non-core discipline for each individual academic.

Finally, we observe that in the macro-areas of Biology, Engineering, Mathematics and Physics, there is no significant association between the character of authorship (diversified/specified) and the number of the authors. In the other four macro-areas the association is significant: positive in Clinical Medicine, and in Earth and Space Sciences; negative in Chemistry and in Biomedical Research.

These first univariate analyses therefore confirm hypotheses H1 and H2. H3 is confirmed overall and in some macro-areas, but in others is refuted.

*Table 3: p-value of one-way hypothesis test for comparison of diversified (Y=1) vs specialized (Y=0) publications (in brackets sign of differences)*

| Macro-area | Multi_discipline vs Mono_discipline | n_discipline | n_authors |
|---|---|---|---|
| Biology | 0.000 (+) | 0.000 (+) | 0.664 (-) |
| Biomedical Research | 0.000 (+) | 0.000 (+) | 0.082 (-) |
| Chemistry | 0.000 (+) | 0.000 (+) | 0.001 (-) |
| Clinical Medicine | 0.000 (+) | 0.000 (+) | 0.000 (+) |
| Earth and Space Sciences | 0.000 (+) | 0.000 (+) | 0.006 (+) |
| Engineering | 0.000 (+) | 0.000 (+) | 0.083 (+) |
| Mathematics | 0.000 (+) | 0.000 (+) | 0.517 (+) |
| Physics | 0.000 (+) | 0.000 (+) | 0.447 (-) |
| Total | 0.000 (+) | 0.000 (+) | 0.000 (+) |

### 3.2 Logistic regression

Multiple regression considering the three independent variables permits precise identification of the effect of each of these on the dependent variable. Given the dichotomous nature of the dependent variable (specialized vs diversified character of publication by a given author), we use a logistic regression.

Table 4 presents the results of this analysis for the overall dataset, showing both the logit coefficient and the odds ratio (OR). The latter represents the odds that Y=1 (the publication is diversified) when X increases by one unit (meaning, for example, the publication is the product of multidisciplinary collaboration with respect to the case that it is not).

*Table 4: Results of the logistic regression for the aggregated dataset*

| Variable | Logit coefficients | Odds ratio | p-value |
|---|---|---|---|
| Multi_discipline | 0.516 | 1.676 | 0.000*** |
| n_discipline | 0.724 | 2.062 | 0.000*** |
| n_authors | -0.180 | 0.835 | 0.000*** |
| Const | -1.359 |  | 0.000*** |

*Dependent variable: 1 for diversified publications; 0 for specialized publications*

*Independent variables: Multi_discipline, dummy variable for presence of co-authors in different core disciplines; n_discipline, number of different core disciplines associated with the authors of a publication; n_authors, number of co-authors of a publication.*

*$*p < 0.05$ $** p < 0.01$ $*** p < 0.001$*

*Prob ($\chi^2$) = 0.000. N=12,604 (co-authorship)*

The value of less than 0.05 for the Prob ($\chi^2$) test indicates that the model has relevant explanatory power. All three independent variables result as statistically significant. The value of the coefficients is positive for the first two variable, with corresponding values of OR greater than 1: the odds ratio for the *Multi_discipline* dummy variable is the factor of the odds that Y=1 (diversified publication) within the *Multi_discipline*=1 category, compared to the odds that Y=1 for the reference (*Multi_discipline*=0) category. Thus, the odds that a publication would be diversified are about 67.6% higher for multi-discipline bylines than for mono-discipline bylines. The increase in the number of core disciplines associated with the co-authors of a paper also corresponds to an increase in the odds for Y=1 (diversified paper), by a factor of 2.062 (+106.2%). Instead, for the *n_authors* variable we see a negative value of Logit coefficient and a corresponding OR less than 1: this implies a negative relationship, meaning that an increase in the number of authors corresponds to lower odds for a publication of being diversified (Y=1), by a factor of -16.5%.

The results are interesting: the fact that the higher the number of disciplines associated with the co-authors, the higher the probability of incurring in research diversification, while the opposite occurs with the number of authors is revealing. It seems to imply that when the research team is too large, the research output not only is published in multidisciplinary journals as expected, but it gives rise also to multiple papers published by a restricted number of co-authors of the original team, in specialized journals. There is then a widespread diffusion of the research output at both multidisciplinary and specialized level.

We repeat the analysis at the level of the macro-areas, for detection of any differences (Table 5). From this, it can be seen that the *Multi_discipline* dummy variable (column 3, Table 5) is not statistically significant in five micro-areas out of eight (Biomedical Research, Clinical Medicine, Earth and Space Sciences, Mathematics and Physics), indicating that in these macro-areas, we cannot affirm with certainty that for a given author, the specialized/diversified character of their publication depends on the fact of it having been achieved in collaboration with colleagues of different core disciplines. For the other macro-areas the values of OR are both significant and greater than 1, in confirmation of hypothesis H2, which was that the presence of multidiscipline co-authors results in an increment in the OR for Y=1, meaning the probability that the publication will be diversified for the author considered.

Referring to the variable *n_discipline* (column 5, Table 5), the relation with the diversified/specialized character of publication is statistically significantly for all macro-areas with the exception of Mathematics and Earth and Space Sciences: the OR is in all cases greater than 1, indicating that the diversified character of a publication authored by a given scientist increases monotonically in probability with the number of different disciplines among the co-authors of that same publication.

Finally, concerning the variable *n_authors*, the relation with the diversified/specialized character is not statistically significant in the macro-areas of Clinical Medicine, Earth and Space Sciences, and Mathematics. In all the other cases the value of odds ratios is significant but less than 1, meaning that an increment in the number of authors corresponds to decreasing odds of being diversified (Y=1).

The logit regression thus confirms hypotheses H1 and H2, both at the overall level and for the individual macro-areas, although in some analyses the result is not significant. Hypothesis H3 is instead refuted at both the overall and macro-area levels.

*Table 5: Odds ratios of the logistic regression, by macro-area*

| Macro-area | Obs | Multi_discipline | | n_discipline | | n_authors | | Prob (χ2) |
|---|---|---|---|---|---|---|---|---|
| Biology | 1,284 | 2.199 | ** | 1.530 | * | 0.832 | *** | 0.000 |
| Biomedical Research | 1,108 | 1.407 | ns | 3.108 | *** | 0.685 | *** | 0.000 |
| Chemistry | 1,746 | 2.296 | *** | 1.964 | *** | 0.651 | *** | 0.000 |
| Clinical Medicine | 1,431 | 1.517 | ns | 2.045 | ** | 1.003 | ns | 0.000 |
| Earth and Space Sciences | 485 | 1.591 | ns | 1.831 | ns | 0.859 | ns | 0.000 |
| Engineering | 4,148 | 1.820 | ** | 2.025 | *** | 0.820 | *** | 0.000 |
| Mathematics | 1,109 | 1.199 | ns | 2.341 | ns | 0.919 | ns | 0.000 |
| Physics | 1,293 | 1.612 | ns | 1.853 | * | 0.853 | * | 0.000 |

*Dependent variable: 1 for diversified publications; 0 for specialized publications*

*Independent variables: Multi_discipline, dummy variable for presence of co-authors in different core disciplines; n_discipline, number of different core disciplines associated with the authors of a publication; n_authors, number of co-authors of a publication.*

*\*p < 0.05; \*\* p < 0.01; \*\*\* p < 0.001; ns= not significant*

## 4. Conclusions

The complexity of modern social and scientific challenges requires integration of knowledge and research collaboration among experts from different disciplines. Many policies at both the national and supranational levels put incentives in play, pushing in this sense.

However the individual researcher faces a difficult choice, since overstepping the boundaries of their own discipline and opening to collaboration with colleagues of other disciplines is risky, difficult and costly. In fact the extreme knowledge complexity produced within each disciplinary area pushes towards ever greater specialization, and incentivizes scientists to collaborate with colleagues of their own discipline.

The observation of incentives which simultaneously push in opposite directions, in evident paradox, should incite scholars of the material to study the phenomenon, and to provide empirical evidence for the design of policies that could favor specialization of knowledge, or its diversification, according to the contextual requirements.

The current work inserts in the stream of studies concerning diversification versus specialization in the scientific activity of individual researchers. Previously, the authors searched for correlations between diversified/specialized research activity and the type of collaboration (intramural, domestic extramural, international) undertaken by individual academics, as observable through co-authorship of scientific publications (Abramo, D'Angelo, & Di Costa, 2017d). In this work the authors have instead examined the influence of the disciplinary composition of the research team on the diversified or specialized character of the participating academic's production.

As would be expected, the analysis confirms that the diversified publications of a scientist are more frequently the fruit of multidisciplinary team collaborations. In addition, the diversified character emerges more evidently with increasing heterogeneity of the team, meaning with increasing number of the core research disciplines associated with the publication co-authors. The result at the overall level is confirmed in examination of all eight of the individual macro-areas. Vice versa, the specialized character in results from a scientific collaboration is linked to the scientist's participation in teams with presence of other researchers from the same core discipline.

Results are relevant at both government and management level. With aims of increasing the pace of scientific and technological advancements, boosting research productivity and addressing complex societal problems, a growing number of national governments and research organizations are making efforts to foster research collaboration, especially multidisciplinary which often underlines research diversification. Shedding light on the relationship between the multidisciplinary character of a research collaboration and diversification/specialization may help formulate coherent policies and synergistic initiatives to actualize them. The increasing adoption of performance based research funding is likely to induce more or less resistance of the research community to the above policies, depending on their impact on research performance of individuals. It has been shown in fact that there is a link between: i) research collaboration and performance of the individual scientist, although the causal nexus between the two has still not been fully clarified (Lee & Bozeman, 2005; He, Geng, & Campbell-Hunt, 2009; Ynalvez & Shrum, 2011; Abramo, D'Angelo, & Murgia, 2017); and ii) research diversification and performance (Abramo, D'Angelo & Di Costa, 2017a). For example, because, all others being equal, the higher the number of authors per publication, the lower the research productivity of individuals (Abramo & D'Angelo, 2014), the finding that the number of co-authors is negatively related to diversification is good news for policies aimed at fostering multidisciplinary research. Alongside, the fact that no differences occur among disciplines, relieve policy makers from the need to differentiate policies across disciplines.

Finally, we warn the practitioner that the interpretation of the results and replication of the analysis at country level require care, given that the results would be sensitive to: i) the convention adopted for definition of specialization/diversification of research activity; ii) the discipline classification scheme for the publications; iii) the specific characteristics of the country system analyzed.

**References**

Abramo, G., & D'Angelo, C. A. (2017a). Gender differences in research diversification behavior. In *Proceedings of the 16th International Society of Scientometrics and Informetrics Conference - (ISSI - 2017), 16-20 October 2017*. Wuhan, China.

Abramo, G., & D'Angelo, C.A., (2017b). Does your surname affect the citability of your publications? *Journal of Informetrics*, 11(1), 121-127.

Abramo, G., D'Angelo, C.A. (2014). How do you define and measure research productivity? *Scientometrics,* 101(2), 1129-1144.

Abramo, G., D'Angelo, C. A., & Cicero, T. (2012). What is the appropriate length of the publication period over which to assess research performance? *Scientometrics*, 93(3), 1005–1017.

Abramo, G., D'Angelo, C. A., & Di Costa, F. (2017a). Diversification vs specialization in research: which strategy pays off? *Working Paper*.

Abramo, G., D'Angelo, C. A., & Di Costa, F. (2017b). Do interdisciplinary research teams deliver higher gains to science? *Scientometrics*, 111(1), 317–336.

Abramo, G., D'Angelo, C. A., & Di Costa, F. (2017c). The effects of gender, age and academic rank on research diversification. *Scientometrics*, 1–15.

Abramo, G., D'Angelo, C. A., & Di Costa, F. (2017d). Authorship analysis of specialized vs diversified research output. *Working Paper*.

Abramo, G., D'Angelo, C.A., & Murgia, G. (2013). The collaboration behaviors of scientists in Italy: A field level analysis. *Journal of Informetrics*, 7(2), 442-454.

Adelman, M. A. (1969). Comment on the "H" concentration measure as a numbers-equivalent. *The Review of Economics and Statistics*, *51*, 99–101.

Boh, W. F., Ren, Y., Kiesler, S., & Bussjaeger, R. (2007). Expertise and collaboration in the geographically dispersed organization. *Organization Science*, 18(4), 595-612.

Bozeman, B., & Corley, E. (2004). Scientists' collaboration strategies: implications for scientific and technical human capital. *Research Policy*, 33(4), 599–616.

Clark, B. Y., & Llorens, J. J. (2012). Investments in scientific research: examining the funding threshold effects on scientific collaboration and variation by academic discipline. *Policy Studies Journal*, 40(4), 698-729.

Cummings, J. N., & Kiesler, S. (2005). Collaborative research across disciplinary and organizational boundaries. Social Studies of Science, 35(5), 703-722.

D'Angelo, C.A., & Abramo, G. (2015). Publication rates in 192 research fields. In A. Salah, Y. Tonta, A.A.A. Salah, C. Sugimoto (Eds) Proceedings of the *15th International Society of Scientometrics and Informetrics Conference - (ISSI - 2015)* (pp. 909-919). Istanbul: Bogazici University Printhouse.

D'Angelo, C. A., Giuffrida, C., & Abramo, G. (2011). A heuristic approach to author name disambiguation in bibliometrics databases for large-scale research assessments. *Journal of the American Society for Information Science and Technology*, 62(2), 257–269.

Darbellay, F. (2015). Rethinking inter- and transdisciplinarity: Undisciplined knowledge and the emergence of a new thought style. *Futures*, 65, 163–174.

He, Z. L., Geng, X. S., & Campbell-Hunt, C. (2009). Research collaboration and research output: A longitudinal study of 65 biomedical scientists in a New Zealand university. *Research Policy*, *38*(2), 306–317.

Jones, B. F., Wuchty, S., & Uzzi, B. (2008). Multi-university research teams: shifting impact, geography, and stratification in science. *Science*, 322(5905), 1259-1262.

Katz, J. S., & Martin, B. R. (1997). What is research collaboration? *Research Policy*, 26(1), 1–18.

Klein, J. T. (2008). Evaluation of interdisciplinary and transdisciplinary research: a literature review. *American journal of preventive medicine*, 35(2), S116-S123.

Lee, S., & Bozeman, B. (2005). The impact of research collaboration on scientific productivity. *Social Studies of Science*, 35(5), 673–702.

Mo, G.Y., (2016). Examining cross-disciplinary communication's impact on multidisciplinary collaborations: Implications for innovations. *Information, Communication & Society,* 19(9), 1250-66.

Mizukami, Y., Mizutani, Y., Honda, K., Suzuki, S., & Nakano, J. (2017). An international research comparative study of the degree of cooperation between disciplines within mathematics and mathematical sciences: proposal and application of new indices for identifying the specialized field of researchers. *Behaviormetrika*, 44(2), 385–403.

Mugabushaka, A.-M., Kyriakou, A., & Papazoglou, T. (2016). Bibliometric indicators of interdisciplinarity: the potential of the Leinster–Cobbold diversity indices to study disciplinary diversity. *Scientometrics*, 107(2), 593–607.

Shimada, Y., & Suzuki, J. (2017). Promoting scientodiversity inspired by biodiversity. *Scientometrics*, 1–17.

Stokols, D., Fuqua, J., Gress, J., Harvey, R., Phillips, K., Baezconde-Garbanati, L., Trochim, W. (2003). Evaluating transdisciplinary science. *Nicotine & Tobacco Research*, 5(6), 21–39.

Strober, M. (2006). Habits of the mind: Challenges for multidisciplinary engagement. *Social Epistemology*, 20(3–4), 315–331.

Viseu, A. (2015). Integration of social science into research is crucial. *Nature, 525*(7569), 291.

Weingart, P. (2000). Interdisciplinarity: The paradoxical discourse. In Peter Weingart and Nico Stehr (Ed.), *Practicing interdisciplinarity* (pp. 25–41). Toronto: University of Toronto Press Inc.

Wray, K. B. (2005). Rethinking scientific specialization. *Social Studies of Science*, 35(1), 151-164.

Wuchty, S., Jones, B. F., & Uzzi, B. (2007). The increasing dominance of teams in production of knowledge. *Science* (New York, N.Y.), 316(5827), 1036–9.

Ynalvez, M. A., & Shrum, W. M. (2011). Professional networks, scientific collaboration, and publication productivity in resource-constrained research institutions in a developing country. *Research Policy*, *40*(2), 204–216